\documentclass[aps,prx,twocolumn,showpacs]{revtex4-1}
\usepackage{bm}
\usepackage{mathrsfs}
\usepackage{amsmath}
\usepackage{amssymb}
\usepackage{graphicx}
\usepackage{amsfonts}
\usepackage{amsthm}
\usepackage{color}
\usepackage{dcolumn}
\usepackage{txfonts}
\usepackage[caption=false]{subfig}
\DeclareMathOperator{\Tr}{Tr}
\DeclareMathOperator{\arccot}{arccot}

\begin{document}

\title{Three-Tangle of a General Three-Qubit State in the Representation of Majorana Stars}
\author{Chon-Fai Kam}
\author{Ren-Bao Liu}
\affiliation{Department of Physics, Centre for Quantum Coherence, and Institute of Theoretical Physics,
The Chinese University of Hong Kong, Shatin, New Territories, Hong Kong, China}

\begin{abstract}
Majorana stars, the $2j$ spin coherent states that are orthogonal to a spin-$j$ state, offer a visualization of general quantum states and may disclose deep structures in quantum states and their evolutions. In particular, the genuine tripartite entanglement - the three-tangle of a symmetric three-qubit state, which can be mapped to a spin-3/2 state, is measured by the normalized product of the distance between the Majorana stars. However, the Majorana representation cannot applied to general non-symmetric $n$-qubit states. We show that after a series of SL$(2,\mathbb{C})$ transformations, non-symmetric three-qubit states can be transformed to symmetric three-qubit states, while at the same time the three-tangle is unchanged. Thus the genuine tripartite entanglement of general three-qubit states has the geometric representation of the associated Majorana stars. The symmetrization and hence the Majorana star representation of certain genuine high-order entanglement for more qubits are possible for some special states. In general cases, however, the constraints on the symmetrization may prevent the Majorana star representation of the genuine entanglement.
\end{abstract}

\maketitle
\section{introduction}\label{I}
In his 1932 seminal paper, Majorana studied the dynamics of a general spin-$j$ state in a time-varying magnetic field, and derived a compact formula for the transition probability \cite{majorana1932atomi}. He generalized Bloch's representation of spin-1/2 state as a single point on unit sphere to a constellation of $2j$ unordered points on unit sphere, known as the Majorana representation. For a general spin-$j$ state expressed in the $|jm\rangle$ basis, $|\psi\rangle=\sum_{m=-j}^jc_m|jm\rangle$, one may introduce a coherent state representation of $|\psi\rangle$ as \cite{perelomov2012generalized}
\begin{equation}
\langle j,\textit{\textbf{n}}|\psi\rangle=\sum_{m=-j}^j\sqrt{\binom{2j}{j+m}}\left(\cos\frac{\theta}{2}\right)^{j-m}\left(\sin\frac{\theta}{2}e^{i\phi}\right)^{j+m}c_m,
\end{equation}
where $\textit{\textbf{n}}\equiv(\theta,\phi)$ and $|j,\textit{\textbf{n}}\rangle$ is the spin-$j$ coherent state \cite{perelomov2012generalized} directed in the direction of $\textit{\textbf{n}}$. Hence, the overlap between the general spin-$j$ state and the spin coherent state which directed in the antipodal direction of $\textit{\textbf{n}}$ is
\begin{subequations}
\begin{align}
\langle j,-\textit{\textbf{n}}|\psi\rangle&=\left(\sin\frac{\theta}{2}\right)^{2j}P(z),\\
P(z)&\equiv \sum_{m=-j}^j(-1)^{j+m}\sqrt{\binom{2j}{j+m}}c_mz^{j+m},
\end{align}
\end{subequations}
where $z\equiv\cot\frac{\theta}{2}e^{i\phi}$ is the stereographic image of $\textit{\textbf{n}}$ from the north pole onto the equatorial plane. Using the fundamental theorem of algebra, the Majorana polynomial $P(z)$ may be factorized as $P(z)=(-1)^{2j}c_j\prod_{k=1}^{2j}(z-z_k)$. Majorana stars are the inverse stereographic image $\textit{\textbf{n}}_k$ of $z_k$. Hence, there are in general $2j$ directions $-\textit{\textbf{n}}_k$ on the unit sphere where $\langle -\textit{\textbf{n}}_k,j|\psi\rangle$ vanishes. In particular, for spin coherent states, the Majorana polynomial has the form $P(z)=(\cos\frac{\theta}{2}-\sin\frac{\theta}{2}e^{-i\phi}z)^{2j}$, which is associated with $2j$ degenerated stars in the direction of $\textit{\textbf{n}}$.

Majorana's representation of general spin states and his transition probability formula were rediscovered several times by Bloch \cite{bloch1945atoms}, Salwen \cite{salwen1955resonance}, Meckler \cite{meckler1958majorana} and Schwinger \cite{schwinger1977majorana}. Majorana's representation was known to mathematicians as ``canonical decomposition" of totally symmetric spinor, and the associated Majorana stars are called ``principal null directions" in spinor theory \cite{penrose1984spinors}. In 1960, Penrose developed a spinor approach to general relativity, and gave an elegant proof of Petrov's classification of gravitational fields based on degeneracy configuration of the ``principal null directions" of gravitational spinor \cite{penrose1960spinor}. Several decades later, after being aware of Majorana's work, Penrose brought it to wider attention via his popular book \cite{penrose1989emperor}. Since then, researches based on Majorana's representation gradually emerged. Zimba and Penrose used Majorana's representation of spin-3/2 state to provide a simplified proof of Bell's non-locality theorem \cite{zimba1993bell}. Inspired by Penrose's works, Hannay studied statistics of Majorana stars for random spin states, and discovered a simple formula for the pair correlation function in the large $j$ limit \cite{hannay1996chaotic}. Two years later, Hannay derived a general formula of Berry's phase for spin states using Majorana's representation \cite{hannay1998berry}, and applied the spin-1 formula to the polarization of light \cite{hannay1998majorana}. Later, Dennis discovered a simple geometric interpretation of polarization singularities in non-paraxial waves in terms of Majorana representation \cite{dennis2001topological}, and gave an alternative proof of Maxwell’s multipole representation of spherical functions using Majorana stars \cite{dennis2004canonical}. 

The visualization of the quantum states as a constellation on unit sphere may be highly valuable in the classification of quantum states and their evolution. It has been used to reveal a beautiful connection between the most sensitive states under small rotations around arbitrary axes in quantum metrology and the platonic solids \cite{kolenderski2008optimal, bouchard2017quantum, chryssomalakos2017optimal, goldberg2018quantum}, and has been applied to classifying novel phases in spinor Bose-Einstein condensates \cite{barnett2006classifying, barnett2007classifying, makela2007inert}.

Remarkably, Majorana's representation of spin states finds application in quantum information science. Bastin \textit{et al.} gave a simple classification of entanglement between symmetric $N$-qubit states under stochastic local operations and classical communication (SLOCC) via degeneracy configuration of the associated Majorana stars \cite{bastin2009operational, mathonet2010entanglement}. In subsequent works, Markham \textit{et al.} \cite{aulbach2010maximally, markham2011entanglement} showed that three types of entanglement measures --- the geometric measure of entanglement, the logarithmic robustness of entanglement, and the relative entropy of entanglement are equivalent when the distribution of Majorana stars obey certain symmetries. Subsequently, Majorana's representation was also used to provide insight into quantum geometric phases and the dynamics of quantum spins \cite{bruno2012quantum, liu2014representation}, and to study the anticoherence of symmetric qubit states \cite{giraud2015tensor, baguette2015anticoherence}. Based on these developments, Majorana's representation of general spin states becomes a valuable tool for visual display of multipartite entanglement between symmetric qubit states.

Entanglement is a resource that is unique to quantum information \cite{wootters1998quantum}, which cannot be increased by local operations when the systems are distributed over spatially separated locations \cite{horodecki2009quantum}. For two-qubit pure states, the entanglement may be measured by Wootters's concurrence $C$ \cite{wootters1998entanglement}, which varies monotonically from 0 to 1 when the state changes from separable to maximally entangled. In particular, for symmetric two-qubit states, which may be written as $|\psi\rangle=|\textit{\textbf{n}}_1\rangle\otimes|\textit{\textbf{n}}_2\rangle+|\textit{\textbf{n}}_2\rangle\otimes|\textit{\textbf{n}}_1\rangle$, Wootters's concurrence becomes $C=\sin^2\frac{\theta_{12}}{2}/(1+\cos^2\frac{\theta_{12}}{2})$ \cite{ribeiro2011entanglement}, where $|\textit{\textbf{n}}_k\rangle\equiv\cos\frac{\theta_k}{2}|0\rangle+\sin\frac
{\theta_k}{2}e^{i\phi_k}|1\rangle$ in the computational basis, and $\theta_{12}\equiv\cos^{-1}(\textit{\textbf{n}}_1\cdot \textit{\textbf{n}}_2)\in[0,\pi]$ is the spherical distance between the Bloch vectors $\textit{\textbf{n}}_1$ and $\textit{\textbf{n}}_2$. The two unordered points $\textit{\textbf{n}}_1$ and $\textit{\textbf{n}}_2$ (Majorana stars) completely determines a symmetric two-qubit state and thus the entanglement. For three-qubit pure states, the entanglement between the parties are measured by 5 independent local unitary invariants: $C_{12}$, $C_{13}$, $C_{23}$, $\kappa$ and $\tau_3$ \cite{kempe1999multiparticle, coffman2000distributed}, where $C_{ij}$ are the pairwise concurrence between the parties $i$ and $j$, $\kappa$ is the Kempe invariant \cite{kempe1999multiparticle}, and $\tau_3$ is the three-tangle, which measures the genuine tripartite entanglement \cite{coffman2000distributed}. In particular, for symmetric three-qubit states, which are written as $|\psi\rangle=\sum_{\sigma\in S_3}|\textit{\textbf{n}}_{\sigma(1)}\rangle\otimes|\textit{\textbf{n}}_{\sigma(2)}\rangle\otimes|\textit{\textbf{n}}_{\sigma(3)}\rangle$, we have $\tau_3=\frac{4}{3}(\prod_{i<j}\sin\frac{\theta_{ij}}{2}/\sum_{i<j}\cos^2\frac{\theta_{ij}}{2})^2$ \cite{ribeiro2011entanglement}, where $S_3$ is the permutation group of order 3. In other words, for symmetric two- and three-qubit pure states, we may measure the genuine entanglement in terms of the distances between the Majorana stars on unit sphere. 

Now, one question naturally arises: can we have a star representation for general two- and three-qubit pure states without permutation symmetries? The benefits of such a representation are evident: it offers an intuitive approach to visualizing the entanglement in terms of three-dimensional geometry; it also provides a simple way to obtain the entanglement --- one just calculates the distances between all Majorana stars on unit sphere. As general non-symmetric states do not possess a Majorana representation, the Majorana star representation cannot be directly employed to formulate entanglement measures. Nevertheless, if we can transform a general non-symmetric pure state to a symmetric state, while at the same time keeping Wootters's concurrence $C$ or the three-tangle $\tau_3$ unchanged, then the Majorana star representation can be applied. For two-qubit states, Schmidt decomposition \cite{bengtsson2017geometry} exists and allows one to express the state as a symmetric state without changing its entanglement properties. Hence, the Majorana star representation for general two-qubit pure states can be immediately obtained from its Schmidt decomposition. As Schmidt decomposition does not exist for three-partite pure states, the Majorana star representation for three-qubit states is not so evident. However, we will show in the following sections that after using Ac\'{i}n's canonical form \cite{acin2000generalized} --- a type of generalized Schmidt decomposition, one may have a star representation of entanglement for general three-qubit states.

The organization of the paper is as follows. In Sec.\:\ref{II}, we will discuss the Majorana star representation of spin states, and the representation of entanglement in terms of Majorana stars. In Sec.\:\ref{III}, we will explicitly construct a set of invertible local transformations $L=L_1\otimes L_2\otimes L_3$, which bring a general non-symmetric three-qubit pure state to a symmetric one, where $L_i\in\mbox{SL}(2,\mathbb{C})$ are special linear transformations of degree 2. As the three-tangle $\tau_3$ is an invariant under special linear transformation \cite{bengtsson2017geometry}, we may express the three-tangle of a general three-qubit state in the constellation of three Majorana stars. In Sec.\:\ref{IV}, we will discuss generalization of such transformation to multi-partite entangled pure states, and will show that similar procedures can be applied to some but not all $n$-qubit states with $n\geq 4$. In Sec.\:\ref{V}, we will discuss mixed entanglement in the Majorana star representation, and will use the mixture of GHZ and W states as an example. Finally, in Sec.\:\ref{VI}, we will discuss the implications and limitations of the current work.

\section{Majorana Representation of Spin States}\label{II}
The essence of the Majorana representation is that a spin-$j$ state can be written as a symmetric tensor product of $N=2j$ spin-1/2 states
\begin{equation}
|\psi\rangle=\frac{1}{\sqrt{N!A_N}}\sum_{\sigma\in S_N}|\textit{\textbf{n}}_{\sigma(1)}\rangle\otimes\cdots\otimes|\textit{\textbf{n}}_{\sigma(N)}\rangle,
\end{equation}
where $A_N\equiv \sum_{\sigma\in S_N}\prod_k\langle \textit{\textbf{n}}_k|\textit{\textbf{n}}_{\sigma(k)}\rangle$ is a normalization factor, $S_N$ is the permutation group of order $N$, and $|\textit{\textbf{n}}_k\rangle$ is a spin-1/2 state polarized along the direction $\textit{\textbf{n}}_k$. The $N=2j$ antipodal directions $-\textit{\textbf{n}}_k$ of the Majorana stars $\textit{\textbf{n}}_k$ are corresponded to the spin-$j$ coherent states $|j,-\textit{\textbf{n}}_k\rangle\equiv |-\textit{\textbf{n}}_k\rangle^{\otimes N}$ that are orthogonal to the spin-$j$ state $|\psi\rangle$. The Majorana representation of spin states can be rephrased as a theorem \cite{penrose1984spinors}: a $2j$-dimensional complex projective space $\mathbf{CP}^{2j}$, which is the state space of a spin-$j$ state, is homeomorphic to a $2j$-fold symmetric tensor product of sphere $\mbox{SP}^{2j}(S^2)$ \cite{liao1954topology, bhatia1983space}, \textit{i.e.}, an ordered tuple $(a_0,a_1,\ldots,a_N)$ in a complex projective space $\mathbf{CP}^N$ is equivalent to an unordered tuple $[\textit{\textbf{n}}_1,\textit{\textbf{n}}_2,\ldots,\textit{\textbf{n}}_N]\equiv\{(\textit{\textbf{n}}_1,\textit{\textbf{n}}_2,\ldots,\textit{\textbf{n}}_N)/\sim|\textit{\textbf{n}}_i\in S^2\}$, where $\sim$ is an equivalence relation defined by $(\textit{\textbf{n}}_1,\textit{\textbf{n}}_2,\ldots,\textit{\textbf{n}}_N)\sim(\textit{\textbf{n}}_{\sigma(1)},\textit{\textbf{n}}_{\sigma(2)},\ldots,\textit{\textbf{n}}_{\sigma(N)})$.

For $N=2$, the Schmidt decomposition of a general two-qubit pure state reads $|\psi\rangle = \mu_1|00\rangle + \mu_2|11\rangle$, where $\mu_1\equiv \cos\chi$ and $\mu_2\equiv \sin\chi$ are the Schmidt coefficients, and $\chi\in[0,\pi/4]$ is the Schmidt angle \cite{acin2000generalized}. The entanglement between the two qubits, measured by Wootters's concurrence $C$, may be written in terms of the Schmidt coefficients: $C\equiv2\mu_1\mu_2=\sin2\chi$ \cite{bengtsson2017geometry}. The entanglement is larger when the Schmidt angle has larger value. As the Schmidt decomposition of a general two-qubit state is already symmetric under permutation of qubits, it can be mapped to a spin-1 state which possesses two Majorana stars with latitudes $\theta_1=\theta_2=\pi-2\arctan\sqrt{\tan\chi}$ and longitudes $\phi_1=\pi/2$ and $\phi_2=3\pi/2$. The Schmidt coefficients and the Wootters concurrence can be expressed via the spherical distance $\theta_{12}=2\theta_1$ between the Majorana stars: $\mu_1=\frac{1}{2}(\sqrt{1+C}+\sqrt{1-C})$, $\mu_2=\frac{1}{2}(\sqrt{1+C}-\sqrt{1-C})$ and $C=\sin^2\frac{\theta_{12}}{2}/(1+\cos^2\frac{\theta_{12}}{2})$. The entanglement between the qubits is larger when the spherical distance between the stars has larger value. Separable states correspond to two identical stars and maximally entangled Bell states correspond to two antipodal stars on equator \cite{ribeiro2011entanglement}.

\section{Representation of Three-tangle using Majorana Stars}\label{III}
As discussed in Sec.\:\ref{I}, the Majorana representation cannot be directly applied to general non-symmetric states. However, we may still find a set of local transformations which send non-symmetric states to symmetric ones without changing the three-tangle, \textit{i.e.}, the global entanglement of three-qubit states. For a general three-qubit state $|\psi_0\rangle=\Gamma_{ijk}|ijk\rangle$, Ac\'{i}n's canonical form, the generalized Schmidt decomposition of three-qubit states reads \cite{acin2000generalized}
\begin{equation}\label{Acin}
    |\psi\rangle=\lambda_0|000\rangle+\lambda_1e^{i\varphi}|100\rangle+\lambda_2|101\rangle+\lambda_3|110\rangle+\lambda_4|111\rangle,
\end{equation}
where $0\leq\varphi\leq \pi$, and $\lambda_0$, $\lambda_1$, $\lambda_2$, $\lambda_3$ and $\lambda_4$ are non-negative real numbers satisfying $\sum_{i=0}^4\lambda_i^2=1$. The relation between Ac\'{i}n's canonical form and the coefficients $\Gamma_{ijk}$ can be specified as follows: let $\mathbf{T}_0$ and $\mathbf{T}_1$ be two matrices with elements $(\mathbf{T}_i)_{jk}\equiv \Gamma_{ijk}$, and let $\mathbf{U}_1$, $\mathbf{U}_2$ and $\mathbf{U}_3$ be three unitary matrices satisfying \cite{acin2000generalized}
\begin{subequations}
\begin{gather}
    \mathbf{T}_i'\equiv \sum_j (\mathbf{U}_1^\dagger)_{ij}\mathbf{T}_j,\:\mbox{s.t.}\:\det \mathbf{T}_0'=0,\\
    \mathbf{U}_2^\dagger\mathbf{T}_0'\mathbf{U}_3=\begin{pmatrix}
    \lambda_0       &  0 \\
    0       &  0
    \end{pmatrix},
    \mathbf{U}^\dagger_2\mathbf{T}_1'\mathbf{U}_3=\begin{pmatrix}
    \lambda_1e^{i\varphi}       &  \lambda_2 \\
    \lambda_3      &  \lambda_4
    \end{pmatrix},
\end{gather}
\end{subequations}
then $|\psi\rangle = \mathbf{U}_1^*\otimes \mathbf{U}_2^*\otimes \mathbf{U}_3|\psi_0\rangle$, where $\mathbf{M}^*$ denotes matrix with complex conjugated entries. $\mathbf{U}_1^*$, $\mathbf{U}_2^*$ and $\mathbf{U}_3$ are unitary matrices, and thus the net transformation $\mathbf{U}_1^*\otimes \mathbf{U}_2^*\otimes \mathbf{U}_3$ is a local unitary. As local unitary transformations do not alter the degree of entanglement, Ac\'{i}n's canonical form preserves entanglement. 

The three-tangle of a general three-qubit state $|\psi_0\rangle$ is proportional to the hyperdeterminant of the third-order tensor $\mathbf{\Gamma}^{(3)}\equiv[\Gamma_{ijk}]$, which may be specified as \cite{bengtsson2017geometry}
\begin{equation} 
\tau_3(|\psi_0\rangle)\equiv 4|\mbox{Det}(\mathbf{\Gamma}^{(3)})|,
\end{equation}
where $\mbox{Det}(\mathbf{\Gamma}^{(3)})$ is Cayley's hyperdeterminant defined by \cite{gelfand2008discriminants}
\begin{gather}
\mbox{Det}(\mathbf{\Gamma}^{(3)})\equiv\left(\begin{vmatrix}
    \Gamma_{000}      & \Gamma_{011}\\
    \Gamma_{100}      & \Gamma_{111}
    \end{vmatrix}+
    \begin{vmatrix}
    \Gamma_{010}      & \Gamma_{001}\\
    \Gamma_{110}      & \Gamma_{101}
    \end{vmatrix}\right)^2\nonumber\\
-4 \begin{vmatrix}
    \Gamma_{000}      & \Gamma_{001}\\
    \Gamma_{100}      & \Gamma_{101}
    \end{vmatrix}\cdot
     \begin{vmatrix}
    \Gamma_{010}      & \Gamma_{011}\\
    \Gamma_{110}      & \Gamma_{111}
    \end{vmatrix}.
\end{gather}
Cayley's hyperdeterminant is a homogeneous polynomial of degree 4. Under invertible local operations $|\tilde{\psi}\rangle=\mathbf{L}_1\otimes \mathbf{L}_2\otimes \mathbf{L}_3|\psi\rangle$, it transforms with a determinantal factor, $\mbox{Det}(\tilde{\mathbf{\Gamma}}^{(3)})=(\det(\mathbf{L}_1))^2(\det(\mathbf{L}_2))^2(\det(\mathbf{L}_3))^2\mbox{Det}(\mathbf{\Gamma}^{(3)})$, where $\mathbf{L}_1$, $\mathbf{L}_2$ and $\mathbf{L}_3$ are invertible matrices \cite{bengtsson2017geometry}. When $\mathbf{L}_1$, $\mathbf{L}_2$ and $\mathbf{L}_3$ are special linear transformations of degree 2, \textit{i.e.}, two-by-two matrices of determinant 1, the hyperdeterminant and the three-tangle become invariants: $\mbox{Det}(\mathbf{\tilde{\Gamma}}^{(3)})=\mbox{Det}(\mathbf{\Gamma}^{(3)})$ and $\tau_3(|\tilde{\psi}\rangle)=\tau_3(|\psi\rangle)$. D{\"u}r \textit{et al.} showed that two states have the same kind of entanglement if both of them can be obtained from the other by means of stochastic local operations and classical communications (SLOCC) \cite{dur2000three}. They proved that two states are equivalent under SLOCC if they are related by invertible local transformations. In other words, the three-tangle $\tau_3$ of general three-qubit states is an SLOCC invariant \cite{dur2000three}.

Using Ac\'{i}n's canonical form, Eq.\:\eqref{Acin}, the three-tangle $\tau_3(|\psi\rangle)$ reads $\tau_3(|\psi\rangle)=4\lambda_0^2\lambda_4^2$. As we are interested in states for which $\tau_3(|\psi\rangle)\neq 0$, we assume $\lambda_0\neq 0$ and $\lambda_4\neq 0$. In order to transform general non-symmetric three-qubit states to symmetric three-qubit states, we consider the following SL$(2,\mathbb{R})$ transformation on the third qubit
\begin{equation}
    \textbf{M}\equiv\begin{pmatrix}
    \gamma       &  0 \\
    g       &  \gamma^{-1}
\end{pmatrix},
g \equiv \frac{\lambda_2\gamma^{-1}-\lambda_3\gamma}{\lambda_4},
\end{equation}
so that after the transformation, the three-qubit state is invariant under permutation of the second and third qubits
\begin{subequations}
\begin{align}
   |\psi'\rangle&\equiv\textbf{I}_2\otimes \textbf{I}_2\otimes \textbf{M}|\psi\rangle\\
   &=\gamma\lambda_0|000\rangle+\frac{\gamma\Delta }{\lambda_4}|100\rangle+\frac{\lambda_2^2+\lambda_4^2}{\lambda_4\gamma}|1nn\rangle,
\end{align}
\end{subequations}
where $\textbf{I}_2$ denotes the two-by-two identity matrix, $|n\rangle\equiv(\lambda_2^2+\lambda_4^2)^{-1/2}(\lambda_2|0\rangle+\lambda_4|1\rangle)$, $\gamma$ is a constant which will be determined later, and $\Delta\equiv \lambda_1\lambda_4e^{i\varphi}-\lambda_2\lambda_3$ vanishes when $|\psi'\rangle$ can be split into two orthogonal product states, $|\psi'\rangle =  \gamma\lambda_0|000\rangle+(\gamma\lambda_4)^{-1}(\lambda_2^2+\lambda_4^2)|1nn\rangle$. In order to proceed further, we consider the following SL$(2,\mathbb{C})$ transformation on the first qubit
\begin{equation}
    \textbf{M}'\equiv\begin{pmatrix}
    a & b \\
    c & d
\end{pmatrix},a\equiv\frac{1}{\lambda_4^2}-\frac{\Delta\lambda_2}{\lambda_0},b\equiv-\frac{\Delta\lambda_4}{\lambda_0},c\equiv\lambda_2\lambda_4,d\equiv\lambda_4^2,
\end{equation}
so that after the transformation, the three-qubit state is invariant under permutation of all the three qubits
\begin{equation}\label{symmetric}
    |\psi''\rangle\equiv \mathbf{M}'\otimes\mathbf{I}_2\otimes\mathbf{I}_2|\psi'\rangle
        =A(|000\rangle+y|nnn\rangle),
\end{equation}
where $A=\gamma\lambda_0\lambda_4^{-2}$ and $y=\gamma^{-2}\lambda_4^2\lambda_0^{-1}(\lambda_2^2+\lambda_4^2)^{3/2}$. Eq.\:\eqref{symmetric} is in Mandilara's canonical form for pure symmetric states \cite{mandilara2014entanglement}, which may be further simplified by performing the following SL$(2,\mathbb{R})$ transformation on all the three qubits
\begin{equation}
    \textbf{M}''\equiv\begin{pmatrix}
    1 & 0 \\
    g' & 1
\end{pmatrix},g'\equiv-\frac{\lambda_2}{\lambda_4},
\end{equation}
so that after the transformation, the three-qubit state has the form
\begin{subequations}
\begin{align}
|\psi'''\rangle&\equiv\mathbf{M}''\otimes\mathbf{M}''\otimes\mathbf{M}''|\psi''\rangle\\
&=\gamma\lambda_0\lambda_4^{-2}|000\rangle+\gamma^{-1}\lambda_4^3|111\rangle.
\end{align}
\end{subequations}
We now fix the parameter $\gamma$ by requiring $\langle\psi'''|\psi'''\rangle=1$. A direct calculation yields $|\psi'''\rangle=\nu_1|000\rangle+\nu_2|111\rangle$, where $\gamma\equiv \lambda_0^{-1}\lambda_4^2\nu_1$ and
\begin{equation*}
\nu_1\equiv(\frac{1}{2}+\frac{1}{2}\sqrt{1-4\lambda_0^2\lambda_4^2})^{1/2}, 
\nu_2\equiv(\frac{1}{2}-\frac{1}{2}\sqrt{1-4\lambda_0^2\lambda_4^2})^{1/2}.
\end{equation*}
As $\nu_1\geq\nu_2$, we may write $|\psi'''\rangle=\cos\vartheta|000\rangle+\sin\vartheta|111\rangle$, which is similar to the Schmidt decomposition of general two-qubit states, where $\vartheta\equiv\cos^{-1}\nu_1\in[0,\pi/4]$. Hence, the three-tangle of the three-qubit state $|\psi'''\rangle$ may be expressed in terms of $\vartheta$ as $\tau_3(|\psi'''\rangle)=\sin^2(2\vartheta)\in[0,1]$, which is an increasing function of $\vartheta$ on $[0,\pi/4]$.

\begin{figure}[tbp]
\subfloat[$|$gGHZ$\rangle$\label{sfig:Fig1}]{%
  \includegraphics[width=0.45\columnwidth]{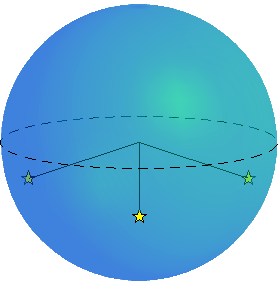}%
}\hfill
\subfloat[$|$gW$\rangle$\label{sfig:Fig2}]{%
  \includegraphics[width=0.45\columnwidth]{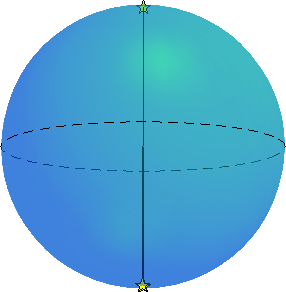}%
}
\caption{The Majorana representation for generalized GHZ states $|$gGHZ$\rangle\equiv a|000\rangle+b|111\rangle$, and generalized W states $|$gW$\rangle\equiv c|001\rangle+d|010\rangle+e|100\rangle$. Here, $a\equiv\cos\vartheta$, $b\equiv \sin\vartheta$, and $\vartheta=\pi/10$.}
\label{MajoranaStarPlot}
\end{figure}

After the transformations $\textbf{M}$, $\textbf{M}'$ and $\textbf{M}''$, Ac\'{i}n's canonical form of three-qubit states becomes a symmetric state, which may be written in terms of the symmetric basic states: $|\psi'''\rangle=\sum_{i=0}^3a_i|S_i^{(3)}\rangle$, where $|S_0^{(3)}\rangle\equiv|000\rangle$, $|S_1^{(3)}\rangle\equiv 3^{-\frac{1}{2}}(|001\rangle+|010\rangle+|100\rangle)$, $|S_2^{(3)}\rangle\equiv 3^{-\frac{1}{2}}(|011\rangle+|101\rangle+|110\rangle)$, and $|S_3^{(3)}\rangle\equiv|111\rangle$. Here, the coefficients $a_i$ are given by $a_0=\cos\vartheta$, $a_1=a_2=0$, and $a_3=\sin\vartheta$. Majorana's star representation may be introduced via the one-to-one correspondence between the symmetric basic states and the conventional $|jm\rangle$ basis states, $\textit{i.e.}$, $|S_{j-m}^{(2j)}\rangle\leftrightarrow|j m\rangle$, so that we have $|\psi'''\rangle=\sum_{m=-j}^jc_m|jm\rangle$ with $c_m= a_{j-m}$ ($j=3/2$). Then the Majorana polynomial of $|\psi'''\rangle$ is obtained via the general formula introduced in Sec.\:\ref{I}
\begin{subequations}
\begin{align}
P(z)&=\sum_{r=0}^{2j}(-1)^{2j-r}\sqrt{C_{2j}^r}a_rz^{2j-r}\\
&=\sin\vartheta-\cos\vartheta z^3.
\end{align}
\end{subequations}
The Majorana polynomial of $|\psi'''\rangle$ has three distinct roots $z_k\equiv (\tan\vartheta)^{1/3}e^{i2k\pi/3}$, ($k=0, 1, 2$). Let us denote the directions of the three Majorana stars on unit sphere as $\mathbf{n}_k$. Hence, $|\psi'''\rangle$ has three Majorana stars distributed evenly on the southern hemisphere with the same latitude $\theta=2\arccot((\tan\vartheta)^{1/3})$ and longitudes $\phi_1=0$, $\phi_2=2\pi/3$, and $\phi_3=4\pi/3$. Then the three-tangle $\tau_3(|\psi'''\rangle)$ can be evaluated by the angles between the directions of the stars \cite{ribeiro2011entanglement}
\begin{equation}\label{MajoranaChordalDistances}
    \tau_3(|\psi'''\rangle)=\frac{4}{3}\left(\frac{\prod_{i<j}\sin\frac{\theta_{ij}}{2}}{\sum_{i<j}\cos^2\frac{\theta_{ij}}{2}}\right)^2=\frac{1}{3}\left(\frac{\prod_{i<j}d_{ij}}{12-\sum_{i<j}d_{ij}^2}\right)^2,
\end{equation}
where $\theta_{ij}\equiv \cos^{-1}(\mathbf{n}_1\cdot \mathbf{n}_2)$ and $d_{ij}\equiv 2\sin\frac{\theta_{ij}}{2}$ are the angle and chordal distances between $\mathbf{n}_i$ and $\mathbf{n}_j$ respectively. As the three Majorana stars of $|\psi'''\rangle$ distributed evenly on the southern hemisphere with the same latitude, the chordal distance between any two Majorana stars are the same, \textit{i.e.}, $d_{12}=d_{13}=d_{23}\equiv d$, which yields $\tau_3(|\psi'''\rangle)=\frac{1}{27}d^6/(4-d^2)^2$, which is an increasing function of $d$ on $[0,2]$. Using elementary geometry, we obtain $d=2\sqrt{3}(R^{-1}+R)^{-1}$, where $R=(\tan\vartheta)^{1/3}$ is the amplitude of the roots $z_k$. A direct calculation yields again $\tau_3(|\psi'''\rangle)=4(R^{-3}+R^3)^{-2}=\sin^2(2\vartheta)$. As the chordal distance between any two Majorana stars of $|\psi'''\rangle$ is an entanglement monotone of genuine tripartite entanglement, the genuine tripartite entanglement is higher when the Majorana stars are closer to the equator -- the closer to the equator, the higher the entanglement. As an example, a generalized GHZ state $|$gGHZ$\rangle\equiv a|000\rangle+b|111\rangle$ with a three-tangle $4a^2b^2$ is already symmetrized, and hence is represented by three distinct Majorana stars distributed evenly on the southern hemisphere with the same latitude (see Fig.\:\ref{sfig:Fig1}).

For a generalized W state $|$gW$\rangle\equiv c|001\rangle+d|010\rangle+e|100\rangle$, the three-tangle vanishes, and hence the above transformations $\textbf{M}$, $\textbf{M}'$ and $\textbf{M}''$ may not be used directly. But we may still apply a set of SL$(2,\mathbb{R})$ transformations to symmetrize it, while keeping the three-tangle unchanged. We may apply the following SL$(2,\mathbb{R})$ transformation on the first qubit
\begin{equation}
\mathbf{T}\equiv\begin{pmatrix}
    \alpha & 0 \\
    0 & \alpha^{-1}
\end{pmatrix},\alpha\equiv\sqrt{\frac{e}{d}},
\end{equation}
so that after the transformation, the generalized W state becomes $|$gW$^\prime\rangle\equiv\mathbf{T}\otimes\textbf{I}_2\otimes \textbf{I}_2|\phi\rangle=c\sqrt{e/d}|001\rangle+\sqrt{de}(|010\rangle+|100\rangle|)$, which is symmetric with respect to the first two qubits. Similarly, we may apply the following SL$(2,\mathbb{R})$ transformation on the third qubit
\begin{equation}
\mathbf{T^\prime}\equiv\begin{pmatrix}
    \beta & 0 \\
    0 & \beta^{-1}
\end{pmatrix},\beta\equiv\sqrt{\frac{c}{d}},
\end{equation}
so that after the transformation, one obtains $|$gW$^{\prime\prime}\rangle\equiv\textbf{I}_2\otimes \textbf{I}_2\otimes\mathbf{T^\prime}|$gW$^\prime\rangle=\sqrt{ce}(|001\rangle+|010\rangle+|100\rangle)$, which is the symmetric W state. The associated Majorana polynomial has three roots $0$ and $\infty$, where $\infty$ is a double root. Hence, a generalized W state is represented by three Majorana stars on unit sphere --- one locates at the south pole and two degenerate ones located at the north pole. It shows that the appearance of a pair of degenerate Majorana stars on the unit sphere indicates the vanishing of three-tangle for a general three-qubit state (see Fig.\:\ref{sfig:Fig2}).

\section{Four-qubit states and beyond}\label{IV}
In the last section, we constructed a series of SL$(2,\mathbb{C})^{\otimes 3}$ transformations which symmetrize a general three-qubit state without changing its genuine tripartite entanglement. We now show that similar procedures can be applied to some but not all $n$-qubit states with $n\geq 4$. 

Let us denote a general four-qubit state as $|\psi\rangle\equiv \Gamma_{ijkl}|ijkl\rangle$. For SLOCC transformation SL$(2,\mathbb{C})^{\otimes 4}$, there exists a set of four independent polynomial invariants. The first polynomial invariant of degree 2 is Cayley's hyperdeterminant defined by \cite{luque2003polynomial}
\begin{align}
H&\equiv \Gamma_{0000}\Gamma_{1111}-\Gamma_{0001}\Gamma_{1110}-\Gamma_{0010}\Gamma_{1101}+\Gamma_{0011}\Gamma_{1100}\nonumber\\
&-\Gamma_{0100}\Gamma_{1011}+\Gamma_{0101}\Gamma_{1010}+\Gamma_{0110}\Gamma_{1001}-\Gamma_{0111}\Gamma_{1000},
\end{align}
and the other two independent polynomial invariants of degree 4 are two determinants given by \cite{luque2003polynomial}
\begin{align}
L&\equiv\begin{vmatrix}
    \Gamma_{0000}      &      \Gamma_{0100}      &     \Gamma_{1000}      &      \Gamma_{1100}\\
    \Gamma_{0001}      &      \Gamma_{0101}      &     \Gamma_{1001}      &      \Gamma_{1101}\\
    \Gamma_{0010}      &      \Gamma_{0110}      &     \Gamma_{1010}      &      \Gamma_{1110}\\
    \Gamma_{0011}      &      \Gamma_{0111}      &      \Gamma_{1011}      &      \Gamma_{1111}\\
    \end{vmatrix},\nonumber\\
M&\equiv\begin{vmatrix}
   \Gamma_{0000}      &      \Gamma_{1000}      &      \Gamma_{0010}      &      \Gamma_{1010}\\
   \Gamma_{0001}       &    \Gamma_{1001}      &      \Gamma_{0011}       &     \Gamma_{1011}\\
   \Gamma_{0100}       &    \Gamma_{1100}       &    \Gamma_{0110}      &      \Gamma_{1110}\\
   \Gamma_{0101}     &      \Gamma_{1101}       &      \Gamma_{0111}     &      \Gamma_{1111}\\
    \end{vmatrix}.
\end{align}
The invariants $L$ and $M$ are closely related to the two-qubit reduced density matrix of the original four-qubit state: $|L|^2=\det\rho_{12}$ and $|M|^2=\det\rho_{13}$, where $\rho_{12}\equiv \Tr_{34}(|\psi\rangle\langle\psi|)$ and $\rho_{13}=\Tr_{24}(|\psi\rangle\langle\psi|)$ \cite{eltschka2012multipartite}. If one denote $\Gamma(\mathbf{x},\mathbf{y},\mathbf{z},\mathbf{t})\equiv\sum_{ijkl}\Gamma_{ijkl}x_iy_jz_kt_l$ and $g_{xy}(\mathbf{x},\mathbf{y})\equiv\det(\partial^2\Gamma/\partial z_i\partial t_j)$, one may define a $3\times 3$ matrix $G_{xy}$ by \cite{luque2003polynomial}
\begin{equation}
g_{xy}(\mathbf{x},\mathbf{y})\equiv 
\begin{pmatrix}
    x_0^2 & x_0x_1 & x_1^2
\end{pmatrix}G_{xy}
\begin{pmatrix}
    y_0^2\\
    y_0y_1\\
    y_1^2
\end{pmatrix},
\end{equation}
For any pair of variables $(\mathbf{u},\mathbf{v})$, one defines $D_{\mu\nu}\equiv\det{(G_{\mu\nu})}$, then the last polynomial invariant of degree 6 is a determinant given by $D\equiv D_{xt}$. In particular, for a four-qubit symmetric state, one obtains $\Gamma_{0001}=\Gamma_{0010}=\Gamma_{0100}=\Gamma_{1000}$, $\Gamma_{0011}=\Gamma_{0101}=\Gamma_{0110}=\Gamma_{1001}=\Gamma_{1010}=\Gamma_{1100}$ and $\Gamma_{0111}=\Gamma_{1011}=\Gamma_{1101}=\Gamma_{1110}$. Hence, one obtains $H=\beta-4\epsilon$, $L=M=0$ and
\begin{equation}
D=\begin{vmatrix}
    \alpha        &       \delta&     \epsilon\\
   \delta      &     \beta      &     \eta\\
      \epsilon    &     \eta     &      \gamma\\
    \end{vmatrix},
\end{equation}
where $\alpha\equiv \Gamma_{0000}\Gamma_{0011}-\Gamma_{0001}^2$, $\beta\equiv \Gamma_{0000}\Gamma_{1111}-\Gamma_{0011}^2$, $\gamma\equiv\Gamma_{0011}\Gamma_{1111}-\Gamma_{0111}^2$, $\delta\equiv\Gamma_{0000}\Gamma_{0111}-\Gamma_{0001}\Gamma_{0011}$, $\epsilon\equiv \Gamma_{0001}\Gamma_{0111}-\Gamma_{0011}^2$ and $\eta\equiv \Gamma_{0001}\Gamma_{1111}-\Gamma_{0011}\Gamma_{0111}$. The vanishing of the invariants $L$ and $M$ for symmetric states can be expected from their unique properties \cite{ren2008permutation}: the invariant $L$ is odd under permutation of the first two qubits (or the last two qubits), and the invariant $M$ is odd under permutation of the first and the third qubits (or the second and the fourth qubits). Hence, both $L$ and $M$ vanish identically for permutation symmetric states.  

Verstraete \cite{verstraete2002four} showed that a general four-qubit state can always be transformed into one of nine distinct SLOCC classes, where each of which is a representative of states interconvertible under SLOCC operations. We may consider the generic class out of the nine SLOCC classes, whose representative is
\begin{gather}
|G_{abcd}\rangle\equiv \frac{a+d}{2}(|0000\rangle+|1111\rangle)+\frac{a-d}{2}(|0011\rangle+|1100\rangle)\nonumber\\
+\frac{b+c}{2}(|0101\rangle+|1010\rangle)+\frac{b-c}{2}(|0110\rangle+|1001\rangle),
\end{gather}
which may also be written as $|G_{abcd}\rangle=a|\Phi_+\Phi_+\rangle+b|\Psi_+\Psi_+\rangle+c|\Psi_-\Psi_-\rangle+d|\Phi_-\Phi_-\rangle$, where $|\Phi_\pm\rangle$ and $|\Psi_\pm\rangle$ are the maximally entangled two-qubit Bell states. The representative of the generic SLOCC class contains the four-qubit GHZ state $|$GHZ$_4\rangle\equiv \frac{1}{\sqrt{2}}(|0000\rangle+|1111\rangle)$ and the EPR pair state $|G_{1000}\rangle=|\Phi_+\Phi_+\rangle$ as special cases. It also contains the four-qubit cluster state \cite{briegel2001persistent} $|\phi_4\rangle=\frac{1}{2}(|0000\rangle+|0011\rangle+|1100\rangle-|1111\rangle)$ as a special case, which can be explicitly obtained from the SL$(2,\mathbb{C})^{\otimes 4}$ transformation
\begin{equation}
|\phi_4\rangle=
\begin{pmatrix}
    e^{-i\pi/8}      & 0\\
    0     & e^{i\pi/8}
    \end{pmatrix}^{\otimes 4}
|G_{abcd}\rangle,
\end{equation}
where $a=(i+e^{-i\pi/4})/2$, $b=c=0$ and $d=(i-e^{-i\pi/4})/2$. The polynomial invariants for the generic SLOCC class are \cite{luque2003polynomial}: $H=\frac{1}{2}(a^2+b^2+c^2+d^2)$, $L=abcd$, $M=[(\frac{c-d}{2})^2-(\frac{a-b}{2})^2][(\frac{a+b}{2})^2-(\frac{c+d}{2})^2]$, and $D=-\frac{1}{4}(ad-bc)(ac-bd)(ab-cd)$. Hence, any states in the generic SLOCC class may be transformed into symmetric states only when the conditions (i): $abcd=0$ and (ii) $c-d=\pm(a-b)$ or $a+b=\pm(c+d)$ are fulfilled. It rules out the possibility that the EPR pair state and four-qubit cluster state can be symmetrized by SL$(2,\mathbb{C})^{\otimes 4}$ transformations. For example, for $c=0$ and $a-d=b$
\begin{align}
|G_{b+d,b,0,d}\rangle&=\frac{b+2d}{2}(|0000\rangle+|1111\rangle)+\frac{b}{2}(|0011\rangle+|1100\rangle\nonumber\\
&+|0101\rangle+|1010\rangle+|0110\rangle+1001\rangle),
\end{align}
where the associated Majorana polynomial has the form $P(z)=\frac{b+2d}{2}z^4+3bz^2+\frac{b+2d}{2}$, which has four roots given by
\begin{equation}
z_k\equiv \pm\sqrt{\frac{-3b\pm\sqrt{9b^2-(b+2d)^2}}{b+2d}}.
\end{equation}
As another example, for $a=1/\sqrt{3}$, $d=\omega/\sqrt{3}$, $b=\omega^2/\sqrt{3}$ and $c=0$, we obtain an entangled four-qubit state which maximizes the average Tsillas-$q$ entropy $E_2^{(q)}$ for $q>2$ \cite{gour2010all}, $|L\rangle\equiv \frac{1}{\sqrt{3}}(|\Phi_+\Phi_+\rangle+\omega|\Phi_-\Phi_-\rangle+\omega^2|\Psi_+\Psi_+\rangle)$, where $\omega\equiv e^{2\pi i/3}$, $E_2^{(q)}\equiv \frac{1}{3}(E^{(q)}_{(AB)(CD)}+E^{(q)}_{(AC)(BD)}+E^{(q)}_{(AD)(BC)})$, $E^{(q)}\equiv \frac{1}{1-q}(\Tr\rho_r^q-1)$ is the Tsallis-$q$ entropy, and $\rho_r\equiv \Tr_B|\psi_{AB}\rangle\langle\psi_{AB}|$ is the reduced density matrix for a given bipartite state $|\psi_{AB}\rangle$. As the conditions $abcd=0$ and $a+b+c+d=0$ are both fulfilled, it is possible to find an SL$(2,\mathbb{C})^{\otimes 4}$ transformation to symmetrize the maximally entangled state $|L\rangle$. Let us consider the following SL$(2,\mathbb{C})^{\otimes 4}$ transformation 
\begin{align}
|L^\prime\rangle&\equiv\mathbf{I}_2\otimes\mathbf{I}_2\otimes\begin{pmatrix}
    0      & 1\\
    -1     & 0
    \end{pmatrix}^{\otimes 2}|L\rangle\nonumber\\
    &= \frac{1-\omega}{\sqrt{3}}(|0000\rangle+|1111\rangle))-\frac{\omega^2}{\sqrt{3}}(|0011\rangle+|1100\rangle\nonumber\\
    &+|0101\rangle+|1010\rangle+|0110\rangle+|1001\rangle),
\end{align}
so that after the transformation, the maximally entangled state is invariant under permutation of qubits. The symmetric state $|L^\prime\rangle$ is associated with a Majorana polynomial $P(z)=\frac{1-\omega}{\sqrt{3}}z^4-\frac{6\omega^2}{\sqrt{3}}z^2+\frac{1-\omega}{\sqrt{3}}$, which has four distinct roots given by $z_k\equiv \pm\sqrt{\frac{3\omega^2\pm 2\sqrt{3\omega}}{1-\omega}}$. As a final example, for the four-qubit GHZ state $|$GHZ$_4\rangle\equiv|G_{\frac{1}{\sqrt{2}}00\frac{1}{\sqrt{2}}}\rangle$, one has $P(z)=(z^4+1)/\sqrt{2}$ and $z_k=e^{i(2k+1)\pi/4}, (k=0,1,2,3)$, which corresponds to four Majorana stars distributed evenly on the equator, with azimuthal angles $\phi_k=(2k+1)\pi/4$ respectively.

Miyake \cite{miyake2003classification} showed that general multi-qubit states under stochastic local operations and classical communication (SLOCC) can be classified by multidimensional determinants \cite{gelfand2008discriminants} similar to Cayley's hyperdeterminant. As multidimensional determinants are invariant under SL$(2,\mathbb{C})^n$ transformation, it seems that we may symmetrize a general $n$-qubit state without changing its global entanglement properties. However, for $n$-qubit states with $n>4$, one may construct polynomial invariants for SLOCC transformation which are odd under permutations of two qubits. For example, for a five-qubit state $|\psi\rangle\equiv \Gamma_{ijklm}|ijklm\rangle$, the unique invariant $F$ of degree 6 can be explicitly written as \cite{dhokovic2009polynomial}
\begin{align}
F=&\Gamma_{i_1j_1k_1l_1m_1}\Gamma_{i_2j_2k_2l_2m_2}\Gamma_{i_3j_3k_3l_3m_3}\Gamma_{i_4j_4k_4l_4m_4}\Gamma_{i_5j_5k_5l_5m_5}\Gamma_{i_6j_6k_6l_6m_6}\nonumber\\
&\epsilon_{k_1k_2}\epsilon_{m_1m_2}\epsilon_{j_1j_3}\epsilon_{l_1l_3}\epsilon_{i_1i_4}\epsilon_{l_2l_4}\epsilon_{m_3m_4}\epsilon_{i_2i_5}\epsilon_{l_5l_6}\epsilon_{m_5m_6}\nonumber\\
&\epsilon_{i_3i_6}\epsilon_{k_3k_6}\epsilon_{j_4j_6}\epsilon_{k_4k_5},
\end{align}
where $\epsilon_{ij}$ is the antisymmetric tensor with $\epsilon_{01}=-\epsilon_{10}=1$ and $\epsilon_{00}=\epsilon_{11}=0$. As the SLOCC invariant $F$ is an odd function under qubit permutations, it vanishes identically for symmetric five-qubit states. Hence, it prevents the symmetrization of general five-qubit states with a non-vanishing value of $F$.

\section{Entangled Multipartite Mixed States}\label{V}
In this section, we briefly discuss the extension of Majorana star representation to multipartite mixed states. The Majorana-like geometric representation of symmetric mixed states, or equivalently mixed spin states, was first introduced by Ramachandran and Ravishankar in 1986 \cite{ramachandran1986polarised}, in which they constructed a set of Majorana stars for the $2j$ Fano statistical tensor parameters which characterize a spin-$j$ assembly. To begin with, one needs the spherical tensor representation of a general spin-$j$ density matrix \cite{fano1957description}
\begin{equation}
\rho = \sum_{k=0}^{2j}\sum_{q=-k}^k\rho_{kq}T_{kq}\equiv\sum_{k=0}^{2j}\boldsymbol{\rho}_k\cdot\boldsymbol{T}_k,
\end{equation}
where $\rho_{kq}\equiv \Tr(\rho T_{kq}^\dagger)$, and $T_{kq}$ are the irreducible tensor operators of rank $k$ in the $2j+1$ dimensional spin space with projection $q$ along the axis of quantization, which can be explicitly expressed in terms of the Clebsch-Gordan coefficients as $T_{kq}\equiv \sum_{m,m'=-j}^j(-1)^{j-m'}C_{jm,j-m'}^{kq}|jm\rangle\langle jm'|$, and satisfy the relations $T_{kq}^\dagger=(-1)^qT_{k-q}$ and $\Tr(T^\dagger_{kq}T_{k'q'})=\delta_{kk'}\delta_{qq'}$. Each vector $\boldsymbol{\rho}_{k}$ is associated with $2k$ Majorana stars defined as the stereographic projection of the roots of the polynomial \cite{suma2017geometric, serrano2019majorana}
\begin{equation}
P^{(k)}(z)\equiv \sum_{q=-k}^k (-1)^{k+q}\sqrt{\binom{2k}{k+q}}\rho_{kq}z^{k+q}.
\end{equation}
The vector $\boldsymbol{\rho}_{0}\equiv\rho_{00}$ does not have an associated constellation of Majorana stars, and its value is fixed to $(2j+1)^{-1}$ by $\Tr\rho=1$. Moreover, since $\rho$ is Hermitian, the condition $T_{kq}^\dagger=(-1)^qT_{k-q}$ implies that $\rho_{kq}^*=(-1)^q\rho_{k-q}$. Hence, the constellation of Majorana stars possesses antipodal symmetry, \textit{i.e.}, $P^{(k)}(z)=(-1)^kz^{2k}(P^{(k)}(-1/z^*))^*$. Unlike that for pure spin states, the constellation for mixed spin states can not fully specify the states, and one needs the relative weights of the irreducible representations. Let us denote $\boldsymbol{\rho}_k=r_k\tilde{\boldsymbol{\rho}}_k$ with respect to a normalized vector $\tilde{\boldsymbol{\rho}}_k$, then the spin-$j$ density matrix $\rho$ may be written as
\begin{equation}
\rho = \frac{\boldsymbol{1}}{2j+1}+\sum_{k=1}^{2j}r_k\tilde{\boldsymbol{\rho}}_k\cdot\boldsymbol{T}_k.
\end{equation}
Hence, any spin-$j$ density matrix $\rho$ is specified by $2j$ spheres with radii $r_k$, and each of which possesses a constellation of $2k$ Majorana stars. 

As a first example, let us consider a spin-1/2 density matrix $\rho=\frac{1}{2}(\boldsymbol{1}+\boldsymbol{r}\cdot\boldsymbol{\sigma})=\frac{\boldsymbol{1}}{2}+\rho_{11}T_{11}+\rho_{10}T_{10}+\rho_{1-1}T_{1-1}$, where $T_{10}=\frac{1}{\sqrt{2}}\sigma_z$, $T_{1 \pm1}=\mp\frac{1}{2}\sigma_{\pm}$ and $\boldsymbol{\rho}_1\equiv(\rho_{11},\rho_{10},\rho_{1-1})=\frac{1}{2}(-r_x+ir_y,\sqrt{2}r_z,r_x+ir_y)$, which yields $r_1=r/\sqrt{2}$. The Majorana polynomial of the vector $\boldsymbol{\rho}_1$ has the form $P^{(1)}(z)=(-r_x+ir_y)z^2-2r_zz+(r_x+ir_y)$, which is associated with a pair of Majorana stars $\boldsymbol{r}$ and $-\boldsymbol{r}$ on a sphere of radius $r/\sqrt{2}$. 

As another example, we consider the density matrix of a symmetric $N$-qubit GHZ state in the $|jm\rangle$ basis, $\rho=|$NGHZ$\rangle\langle$NGHZ$|$, where $N=2j$ and $|$NGHZ$\rangle\equiv \frac{1}{\sqrt{2}}(|jj\rangle+|j-j\rangle)$. As the only non-zero matrix elements of $\rho$ is $\rho_{jj}=\rho_{j-j}=\rho_{-jj}=\rho_{-j-j}=\frac{1}{2}$, we obtain $\rho_{k0}=\frac{1}{2}[1+(-1)^k]C^{k0}_{jjj-j}$ and $\rho_{2j-2j}=(-1)^{2j}\rho_{2j2j}=\frac{1}{2}(-1)^{2j}C_{jjjj}^{2j2j}$. For $k<2j$, the Majorana polynomial for the vector $\boldsymbol{\rho}_k$ is $P^{(k)}(z)=(-1)^k\sqrt{\binom{2k}{k}}\rho_{k0}z^k$, which has a multiple root $0$ of multiplicity $k$ for $k$ even, and has a multiple root $\infty$ of multiplicity $k$ for $k$ odd. Hence, for $k<2j$, there are $k$ degenerate Majorana stars at the north pole for $k$ even, while there are $k$ degenerate Majorana stars at the south pole for $k$ odd. In contrast, for $k=2j$, the Majorana polynomial has the form $P^{(2j)}(z)=\frac{1}{2}z^{4j}+\frac{1}{2}(1+(-1)^{2j})z^{2j}+\frac{1}{2}(-1)^{2j}$, which has $4j$ distinct roots $e^{2\pi in/(4j)}$ with $n=0, 1, ..., 4j-1$ for $2j$ odd, and has $2j$ double roots $e^{\pi i m/(2j)}$ with $m=0,1, ..., 2j-1$ for $2j$ even.

\section{Conclusion}\label{VI}
We have found a way to visually represent the genuine tripartite entanglement of general three-qubit pure states. We used Ac\'{i}n's canonical form of general three-qubits states, and transformed it into a symmetric form similar to the Schmidt decomposition of general two-qubit states via a series of SLOCC transformations which keep the three-tangle invariant. Based on Majorana's representation of spin states, we projected the symmetrized state onto a coherent state, and obtained a set of three Majorana stars on unit sphere which distributes evenly on the southern hemisphere with the same latitude. The genuine tripartite entanglement is then visually represented by the chordal distance between any two Majorana stars. Such a representation may become a useful tool in the field of quantum computation and information.

Although our work is limited to the representation of genuine tripartite entanglement of three-qubit states, the current approach can be applied to some important four-qubit states, including the four-qubit GHZ state and the entangled four-qubit state which maximizes the average Tsallis-$q$ entropy for $q>2$. However, due to the fact that there exist multi-partite entangled states which cannot be symmetrized by SLOCC transformations, the Majorana representation can only be applied to some but not all $n$-qubit entangled states with $n\geq 4$.
 
\begin{acknowledgements}
This work was supported by Hong Kong RGC/GRF Project 14304117.
\end{acknowledgements}


\begin{thebibliography}{10}%
\makeatletter
\providecommand \@ifxundefined [1]{%
 \ifx #1\undefined \expandafter \@firstoftwo
 \else \expandafter \@secondoftwo
\fi
}%
\providecommand \@ifnum [1]{%
 \ifnum #1\expandafter \@firstoftwo
 \else \expandafter \@secondoftwo
\fi
}%
\providecommand \enquote [1]{``#1''}%
\providecommand \bibnamefont  [1]{#1}%
\providecommand \bibfnamefont [1]{#1}%
\providecommand \citenamefont [1]{#1}%
\providecommand\href[0]{\@sanitize\@href}%
\providecommand\@href[1]{\endgroup\@@startlink{#1}\endgroup\@@href}%
\providecommand\@@href[1]{#1\@@endlink}%
\providecommand \@sanitize [0]{\begingroup\catcode`\&12\catcode`\#12\relax}%
\@ifxundefined \pdfoutput {\@firstoftwo}{%
 \@ifnum{\z@=\pdfoutput}{\@firstoftwo}{\@secondoftwo}%
}{%
 \providecommand\@@startlink[1]{\leavevmode}%
 \providecommand\@@endlink[0]{}%
}{%
 \providecommand\@@startlink[1]{%
  \leavevmode
  \pdfstartlink
   attr{/Border[0 0 1 ]/H/I/C[0 1 1]}%
   user{/Subtype/Link/A<</Type/Action/S/URI/URI(#1)>>}%
  \relax
 }%
 \providecommand\@@endlink[0]{\pdfendlink}%
}%
\providecommand \url  [0]{\begingroup\@sanitize \@url }%
\providecommand \@url [1]{\endgroup\@href {#1}{\urlprefix}}%
\providecommand \urlprefix [0]{URL }%
\providecommand \Eprint[0]{\href }%
\@ifxundefined \urlstyle {%
  \providecommand \doi [1]{doi:\discretionary{}{}{}#1}%
}{%
  \providecommand \doi [0]{doi:\discretionary{}{}{}\begingroup
  \urlstyle{rm}\Url }%
}%
\providecommand \doibase [0]{http://dx.doi.org/}%
\providecommand \Doi[1]{\href{\doibase#1}}%
\providecommand \bibAnnote [3]{%
  \BibitemShut{#1}%
  \begin{quotation}\noindent
    \textsc{Key:}\ #2\\\textsc{Annotation:}\ #3%
  \end{quotation}%
}%
\providecommand \bibAnnoteFile [2]{%
  \IfFileExists{#2}{\bibAnnote {#1} {#2} {\input{#2}}}{}%
}%
\providecommand \typeout [0]{\immediate \write \m@ne }%
\providecommand \selectlanguage [0]{\@gobble}%
\providecommand \bibinfo [0]{\@secondoftwo}%
\providecommand \bibfield [0]{\@secondoftwo}%
\providecommand \translation [1]{[#1]}%
\providecommand \BibitemOpen[0]{}%
\providecommand \bibitemStop [0]{}%
\providecommand \bibitemNoStop [0]{.\EOS\space}%
\providecommand \EOS [0]{\spacefactor3000\relax}%
\providecommand \BibitemShut [1]{\csname bibitem#1\endcsname}%

\bibitem{majorana1932atomi}%
  \BibitemOpen
  \bibfield{author}{%
  \bibinfo {author} {\bibfnamefont{E.}~\bibnamefont{Majorana}},\ }%
  \bibfield{journal}{%
  \bibinfo {journal} {Il Nuo. Cim.}\ }%
  \textbf{\bibinfo {volume} {9}},\ \bibinfo {pages} {43} (\bibinfo {year}
  {1932}).%
  \bibAnnoteFile{NoStop}{majorana1932atomi}%
  
\bibitem{perelomov2012generalized}%
  \BibitemOpen
  \bibfield{author}{%
  \bibinfo {author} {\bibfnamefont{A.}~\bibnamefont{Perelomov}},\ }%
  \bibinfo {title} {\textit{Generalized coherent states and their applications} (Springer Science \& Business Media, 2012)}.%
  \bibAnnoteFile{NoStop}{perelomov2012generalized}%
  
 \bibitem{bloch1945atoms}%
  \BibitemOpen
  \bibfield{author}{%
  \bibinfo {author} {\bibfnamefont{F.}~\bibnamefont{Bloch}}\ and\ 
  \bibinfo {author} {\bibfnamefont{I. I.}\ \bibnamefont{Rabi}},\ }%
  \bibfield{journal}{%
  \bibinfo {journal} {Rev. Mod. Phy.}\ }%
  \textbf{\bibinfo {volume} {17}},\ \bibinfo {pages} {237} (\bibinfo {year}
  {1945}).%
 \bibAnnoteFile{NoStop}{bloch1945atoms}%
 
 \bibitem{salwen1955resonance}%
  \BibitemOpen
  \bibfield{author}{%
  \bibinfo {author} {\bibfnamefont{A.}~\bibnamefont{Meckler}},\ }%
  \bibfield{journal}{%
  \bibinfo {journal} {Phys. Rev.}\ }%
  \textbf{\bibinfo {volume} {99}},\ \bibinfo {pages} {1274} (\bibinfo {year}
  {1955}).%
  \bibAnnoteFile{NoStop}{salwen1955resonance}%
 
 \bibitem{meckler1958majorana}%
  \BibitemOpen
  \bibfield{author}{%
  \bibinfo {author} {\bibfnamefont{H.}~\bibnamefont{Salwen}},\ }%
  \bibfield{journal}{%
  \bibinfo {journal} {Phys. Rev.}\ }%
  \textbf{\bibinfo {volume} {111}},\ \bibinfo {pages} {1447} (\bibinfo {year}
  {1958}).%
  \bibAnnoteFile{NoStop}{meckler1958majorana}%
 
\bibitem{schwinger1977majorana}%
  \BibitemOpen
  \bibfield{author}{%
  \bibinfo {author} {\bibfnamefont{J.}~\bibnamefont{Schwinger}},\ }%
  \bibfield{journal}{%
  \bibinfo {journal} {Trans. N. Y. Acad. Sci.}\ }%
  \textbf{\bibinfo {volume} {38}},\ \bibinfo {pages} {170} (\bibinfo {year}
  {1977}).%
  \bibAnnoteFile{NoStop}{schwinger1977majorana}%
  
\bibitem{penrose1984spinors}%
  \BibitemOpen
  \bibfield{author}{%
    \bibinfo {author} {\bibfnamefont{R.}~\bibnamefont{Penrose}}\ and\
  \bibinfo {author} {\bibfnamefont{W.}\ \bibnamefont{Rindler}},\ }%
  \bibinfo {title} {\textit{Spinors and space-time} (Cambridge University Press, 1984), Vol. 1}.%
  \bibAnnoteFile{NoStop}{penrose1984spinors}%
  
\bibitem{penrose1960spinor}%
  \BibitemOpen
  \bibfield{author}{%
  \bibinfo {author} {\bibfnamefont{R.}~\bibnamefont{Penrose}},\ }%
  \bibfield{journal}{%
  \bibinfo {journal} {Ann. Phys.}\ }%
  \textbf{\bibinfo {volume} {10}},\ \bibinfo {pages} {171} (\bibinfo {year}
  {1960}).%
  \bibAnnoteFile{NoStop}{penrose1960spinor}%
  
\bibitem{penrose1989emperor}%
  \BibitemOpen
  \bibfield{author}{%
    \bibinfo {author} {\bibfnamefont{R.}~\bibnamefont{Penrose}},\ }%
  \bibinfo {title} {\textit{The Emperor's New Mind} (Oxford University Press, 1989)}.%
  \bibAnnoteFile{NoStop}{penrose1989emperor}%

 \bibitem{zimba1993bell}%
  \BibitemOpen
  \bibfield{author}{%
  \bibinfo {author} {\bibfnamefont{J.}~\bibnamefont{Zimba}}\ and\ 
  \bibinfo {author} {\bibfnamefont{R.}\ \bibnamefont{Penrose}},\ }%
  \bibfield{journal}{%
  \bibinfo {journal} {Stud. Hist. Phil. Sci. A}\ }%
  \textbf{\bibinfo {volume} {24}},\ \bibinfo {pages} {697} (\bibinfo {year}
  {1993}).%
 \bibAnnoteFile{NoStop}{zimba1993bell}%

\bibitem{hannay1996chaotic}%
  \BibitemOpen
  \bibfield{author}{%
  \bibinfo {author} {\bibfnamefont{J. H.}~\bibnamefont{Hannay}},\ }%
  \bibfield{journal}{%
  \bibinfo {journal} {J. Phys. A: Math. Theor.}\ }%
  \textbf{\bibinfo {volume} {29}},\ \bibinfo {pages} {L101} (\bibinfo {year}
  {1996}).%
  \bibAnnoteFile{NoStop}{hannay1996chaotic}%
  
  \bibitem{hannay1998berry}%
  \BibitemOpen
  \bibfield{author}{%
  \bibinfo {author} {\bibfnamefont{J. H.}~\bibnamefont{Hannay}},\ }%
  \bibfield{journal}{%
  \bibinfo {journal} {J. Phys. A: Math. Theor.}\ }%
  \textbf{\bibinfo {volume} {31}},\ \bibinfo {pages} {L53} (\bibinfo {year}
  {1998}).%
  \bibAnnoteFile{NoStop}{hannay1998berry}%
  
\bibitem{hannay1998majorana}%
  \BibitemOpen
  \bibfield{author}{%
  \bibinfo {author} {\bibfnamefont{J. H.}~\bibnamefont{Hannay}},\ }%
  \bibfield{journal}{%
  \bibinfo {journal} {J. Mod. Opt.}\ }%
  \textbf{\bibinfo {volume} {45}},\ \bibinfo {pages} {1001} (\bibinfo {year}
  {1998}).%
  \bibAnnoteFile{NoStop}{hannay1998majorana}%
  
\bibitem{dennis2001topological}%
  \BibitemOpen
  \bibfield{author}{%
  \bibinfo {author} {\bibfnamefont{M. R.}~\bibnamefont{Dennis}},\ }%
  \bibinfo {title} {\textit{Topological singularities in wave fields} (PhD Thesis, University of Bristol, 2001)}.%
  \bibAnnoteFile{NoStop}{dennis2001topological}%
  
\bibitem{dennis2004canonical}%
  \BibitemOpen
  \bibfield{author}{%
  \bibinfo {author} {\bibfnamefont{M. R.}~\bibnamefont{Dennis}},\ }%
\bibfield{journal}{%
  \bibinfo {journal} {J. Phys. A: Math. Theor.}\ }%
  \textbf{\bibinfo {volume} {37}},\ \bibinfo {pages} {9487} (\bibinfo {year}
  {2004}).%
 \bibAnnoteFile{NoStop}{dennis2004canonical}%

 \bibitem{kolenderski2008optimal}%
  \BibitemOpen
  \bibfield{author}{%
  \bibinfo {author} {\bibfnamefont{P.}~\bibnamefont{Kolenderski}}\ and\ 
  \bibinfo {author} {\bibfnamefont{R.}\ \bibnamefont{Demkowicz-Dobrzanski}},\ }%
  \bibfield{journal}{%
  \bibinfo {journal} {Phys. Rev. A}\ }%
  \textbf{\bibinfo {volume} {78}},\ \bibinfo {pages} {052333} (\bibinfo {year}
  {2008}).%
 \bibAnnoteFile{NoStop}{kolenderski2008optimal}%
 
\bibitem{bouchard2017quantum}%
  \BibitemOpen
  \bibfield{author}{%
  \bibinfo {author} {\bibfnamefont{F.}~\bibnamefont{Bouchard}},
  \bibinfo {author} {\bibfnamefont{P.}~\bibnamefont{de la Hoz}},
  \bibinfo {author} {\bibfnamefont{G.}\ \bibnamefont{Bj{\"o}rk}},
  \bibinfo {author} {\bibfnamefont{R. W.}\ \bibnamefont{Boyd}},
  \bibinfo {author} {\bibfnamefont{M.}\ \bibnamefont{Grassl}},
  \bibinfo {author} {\bibfnamefont{Z.}\ \bibnamefont{Hradil}},
   \bibinfo {author} {\bibfnamefont{E.}\ \bibnamefont{Karimi}},
  \bibinfo {author} {\bibfnamefont{A. B.}\ \bibnamefont{Klimov}},
     \bibinfo {author} {\bibfnamefont{G.}\ \bibnamefont{Leuchs}},
  \bibinfo {author} {\bibfnamefont{J.}\ \bibnamefont{{\v{R}}eh{\'a}{\v{c}}ek}}, \ and\ 
  \bibinfo {author} {\bibfnamefont{L. L.}\ \bibnamefont{S{\'a}nchez-Soto}},\ }%
  \bibfield{journal}{%
  \bibinfo {journal} {Optica}\ }%
  \textbf{\bibinfo {volume} {4}},\ \bibinfo {pages} {1429} (\bibinfo {year}
  {2017}).%
 \bibAnnoteFile{NoStop}{bouchard2017quantum}%

 \bibitem{chryssomalakos2017optimal}%
  \BibitemOpen
  \bibfield{author}{%
  \bibinfo {author} {\bibfnamefont{C.}~\bibnamefont{Chryssomalakos}}\ and\ 
  \bibinfo {author} {\bibfnamefont{H.}\ \bibnamefont{Hern{\'a}ndez-Coronado}},\ }%
  \bibfield{journal}{%
  \bibinfo {journal} {Phys. Rev. A}\ }%
  \textbf{\bibinfo {volume} {95}},\ \bibinfo {pages} {052125} (\bibinfo {year}
  {2017}).%
 \bibAnnoteFile{NoStop}{chryssomalakos2017optimal}%

 \bibitem{goldberg2018quantum}%
  \BibitemOpen
  \bibfield{author}{%
  \bibinfo {author} {\bibfnamefont{A. C.}~\bibnamefont{Goldberg}}\ and\ 
  \bibinfo {author} {\bibfnamefont{D. F. V.}\ \bibnamefont{James}},\ }%
  \bibfield{journal}{%
  \bibinfo {journal} {Phys. Rev. A}\ }%
  \textbf{\bibinfo {volume} {98}},\ \bibinfo {pages} {032113} (\bibinfo {year}
  {2018}).%
 \bibAnnoteFile{NoStop}{goldberg2018quantum}%

 \bibitem{barnett2006classifying}%
  \BibitemOpen
  \bibfield{author}{%
  \bibinfo {author} {\bibfnamefont{R.}~\bibnamefont{Barnett}},
  \bibinfo {author} {\bibfnamefont{A.}\ \bibnamefont{Turner}},\ and\ \bibinfo
  {author} {\bibfnamefont{E.}\ \bibnamefont{Demler}},\ }%
  \bibfield{journal}{%
  \bibinfo {journal} {Phys. Rev. Lett.}\ }%
  \textbf{\bibinfo {volume} {97}},\ \bibinfo {pages} {180412} (\bibinfo {year}
  {2006}).%
 \bibAnnoteFile{NoStop}{barnett2006classifying}%

 \bibitem{barnett2007classifying}%
  \BibitemOpen
  \bibfield{author}{%
  \bibinfo {author} {\bibfnamefont{R.}~\bibnamefont{Barnett}},
  \bibinfo {author} {\bibfnamefont{A.}\ \bibnamefont{Turner}},\ and\ \bibinfo
  {author} {\bibfnamefont{E.}\ \bibnamefont{Demler}},\ }%
  \bibfield{journal}{%
  \bibinfo {journal} {Phys. Rev. A}\ }%
  \textbf{\bibinfo {volume} {76}},\ \bibinfo {pages} {013605} (\bibinfo {year}
  {2007}).%
 \bibAnnoteFile{NoStop}{barnett2007classifying}%

 \bibitem{makela2007inert}%
  \BibitemOpen
  \bibfield{author}{%
  \bibinfo {author} {\bibfnamefont{H.}~\bibnamefont{M{\"a}kel{\"a}}}\ and\ 
  \bibinfo {author} {\bibfnamefont{K. A.}\ \bibnamefont{Suominen}},\ }%
  \bibfield{journal}{%
  \bibinfo {journal} {Phys. 99. Lett.}\ }%
  \textbf{\bibinfo {volume} {95}},\ \bibinfo {pages} {190408} (\bibinfo {year}
  {2007}).%
 \bibAnnoteFile{NoStop}{makela2007inert}%
 
\bibitem{bastin2009operational}%
  \BibitemOpen
  \bibfield{author}{%
  \bibinfo {author} {\bibfnamefont{T.}~\bibnamefont{Bastin}},
  \bibinfo {author} {\bibfnamefont{S.}~\bibnamefont{Krins}},
  \bibinfo {author} {\bibfnamefont{P.}\ \bibnamefont{Mathonet}},
  \bibinfo {author} {\bibfnamefont{M.}\ \bibnamefont{Godefroid}},
  \bibinfo {author} {\bibfnamefont{L.}\ \bibnamefont{Lamata}}, \ and\ 
  \bibinfo {author} {\bibfnamefont{E.}\ \bibnamefont{Solano}},\ }%
  \bibfield{journal}{%
  \bibinfo {journal} {Phys. Rev. Lett.}\ }%
  \textbf{\bibinfo {volume} {103}},\ \bibinfo {pages} {070503} (\bibinfo {year}
  {2009}).%
 \bibAnnoteFile{NoStop}{bastin2009operational}%
 
 \bibitem{mathonet2010entanglement}%
  \BibitemOpen
  \bibfield{author}{%
  \bibinfo {author} {\bibfnamefont{P.}~\bibnamefont{Mathonet}},
  \bibinfo {author} {\bibfnamefont{S.}~\bibnamefont{Krins}},
  \bibinfo {author} {\bibfnamefont{M.}\ \bibnamefont{Godefroid}},
  \bibinfo {author} {\bibfnamefont{L.}\ \bibnamefont{Lamata}},
  \bibinfo {author} {\bibfnamefont{E.}\ \bibnamefont{Solano}}, \ and\ 
  \bibinfo {author} {\bibfnamefont{T.}~\bibnamefont{Bastin}},\ }%
  \bibfield{journal}{%
  \bibinfo {journal} {Phys. Rev. A}\ }%
  \textbf{\bibinfo {volume} {81}},\ \bibinfo {pages} {052315} (\bibinfo {year}
  {2010}).%
 \bibAnnoteFile{NoStop}{mathonet2010entanglement}%
 
 \bibitem{aulbach2010maximally}%
  \BibitemOpen
  \bibfield{author}{%
  \bibinfo {author} {\bibfnamefont{M.}~\bibnamefont{Aulbach}},
  \bibinfo {author} {\bibfnamefont{D.}~\bibnamefont{Markham}},\ and\ 
  \bibinfo {author} {\bibfnamefont{M.}\ \bibnamefont{Murao}},\ }%
  \bibfield{journal}{%
  \bibinfo {journal} {New. J. Phys.}\ }%
  \textbf{\bibinfo {volume} {12}},\ \bibinfo {pages} {073025} (\bibinfo {year}
  {2010}).%
 \bibAnnoteFile{NoStop}{aulbach2010maximally}%
 
 \bibitem{markham2011entanglement}%
  \BibitemOpen
  \bibfield{author}{%
  \bibinfo {author} {\bibfnamefont{D. J. H.}~\bibnamefont{Markham}},\ }%
  \bibfield{journal}{%
  \bibinfo {journal} {Phys. Rev. A}\ }%
  \textbf{\bibinfo {volume} {83}},\ \bibinfo {pages} {042332} (\bibinfo {year}
  {2011}).%
  \bibAnnoteFile{NoStop}{markham2011entanglement}%
  
 \bibitem{bruno2012quantum}%
  \BibitemOpen
  \bibfield{author}{%
  \bibinfo {author} {\bibfnamefont{P.}~\bibnamefont{Bruno}},\ }%
\bibfield{journal}{%
  \bibinfo {journal} {Phys. Rev. Lett.}\ }%
  \textbf{\bibinfo {volume} {108}},\ \bibinfo {pages} {240402} (\bibinfo {year}
  {2012}).%
 \bibAnnoteFile{NoStop}{bruno2012quantum}%
 
  \bibitem{liu2014representation}%
  \BibitemOpen
  \bibfield{author}{%
  \bibinfo {author} {\bibfnamefont{H. D.}~\bibnamefont{Liu}}\ and\ 
  \bibinfo {author} {\bibfnamefont{L. B.}\ \bibnamefont{Fu}},\ }%
  \bibfield{journal}{%
  \bibinfo {journal} {Phys. Rev. Lett.}\ }%
  \textbf{\bibinfo {volume} {113}},\ \bibinfo {pages} {240403} (\bibinfo {year}
  {2014}).%
 \bibAnnoteFile{NoStop}{liu2014representation}%
 
 \bibitem{giraud2015tensor}%
  \BibitemOpen
  \bibfield{author}{%
  \bibinfo {author} {\bibfnamefont{O.}~\bibnamefont{Giraud}},
  \bibinfo {author} {\bibfnamefont{D.}~\bibnamefont{Braun}},
    \bibinfo {author} {\bibfnamefont{D.}~\bibnamefont{Baguette}},
  \bibinfo {author} {\bibfnamefont{T.}\ \bibnamefont{Bastin}},\ and\ 
  \bibinfo {author} {\bibfnamefont{J.}\ \bibnamefont{Martin}},\ }%
  \bibfield{journal}{%
  \bibinfo {journal} {Phys. Rev. Lett.}\ }%
  \textbf{\bibinfo {volume} {114}},\ \bibinfo {pages} {080401} (\bibinfo {year}
  {2015}).%
 \bibAnnoteFile{NoStop}{giraud2015tensor}%
 
  \bibitem{baguette2015anticoherence}%
  \BibitemOpen
  \bibfield{author}{%
  \bibinfo {author} {\bibfnamefont{D.}~\bibnamefont{Baguette}},
  \bibinfo {author} {\bibfnamefont{F.}~\bibnamefont{Damanet}},
  \bibinfo {author} {\bibfnamefont{O.}\ \bibnamefont{Giraud}},\ and\ 
  \bibinfo {author} {\bibfnamefont{J.}\ \bibnamefont{Martin}},\ }%
  \bibfield{journal}{%
  \bibinfo {journal} {Phys. Rev. A}\ }%
  \textbf{\bibinfo {volume} {92}},\ \bibinfo {pages} {052333} (\bibinfo {year}
  {2015}).%
 \bibAnnoteFile{NoStop}{giraud2015tensor}%

\bibitem{wootters1998quantum}%
  \BibitemOpen
  \bibfield{author}{%
  \bibinfo {author} {\bibfnamefont{W. K.}~\bibnamefont{Wootters}},\ }%
  \bibfield{journal}{%
  \bibinfo {journal} {Phil. Trans. R. Soc. Lond. A}\ }%
  \textbf{\bibinfo {volume} {356}},\ \bibinfo {pages} {1717} (\bibinfo {year}
  {1998}).%
  \bibAnnoteFile{NoStop}{wootters1998quantum}%
  
 \bibitem{horodecki2009quantum}%
  \BibitemOpen
  \bibfield{author}{%
  \bibinfo {author} {\bibfnamefont{R.}~\bibnamefont{Horodecki}},
  \bibinfo {author} {\bibfnamefont{P.}~\bibnamefont{Horodecki}},
  \bibinfo {author} {\bibfnamefont{M.}\ \bibnamefont{Horodecki}},\ and\ 
  \bibinfo {author} {\bibfnamefont{K.}\ \bibnamefont{Horodecki}},\ }%
  \bibfield{journal}{%
  \bibinfo {journal} {Rev. Mod. Phys.}\ }%
  \textbf{\bibinfo {volume} {81}},\ \bibinfo {pages} {865} (\bibinfo {year}
  {2009}).%
 \bibAnnoteFile{NoStop}{horodecki2009quantum}%
  
\bibitem{wootters1998entanglement}%
  \BibitemOpen
  \bibfield{author}{%
  \bibinfo {author} {\bibfnamefont{W. K.}~\bibnamefont{Wootters}},\ }%
  \bibfield{journal}{%
  \bibinfo {journal} {Phys. Rev. Lett.}\ }%
  \textbf{\bibinfo {volume} {80}},\ \bibinfo {pages} {2245} (\bibinfo {year}
  {1998}).%
  \bibAnnoteFile{NoStop}{wootters1998entanglement}%

\bibitem{ribeiro2011entanglement}%
  \BibitemOpen
  \bibfield{author}{%
  \bibinfo {author} {\bibfnamefont{P.}~\bibnamefont{Ribeiro}}\ and\
  \bibinfo {author} {\bibfnamefont{R.}\ \bibnamefont{Mosseri}},\ }%
  \bibfield{journal}{%
  \bibinfo {journal} {Phys. Rev. Lett.}\ }%
  \textbf{\bibinfo {volume} {106}},\ \bibinfo {pages} {180502} 
  (\bibinfo {year}{2011}).%
 \bibAnnoteFile{NoStop}{ribeiro2011entanglement}%

\bibitem{kempe1999multiparticle}%
  \BibitemOpen
  \bibfield{author}{%
  \bibinfo {author} {\bibfnamefont{J.}~\bibnamefont{Kempe}},\ }%
  \bibfield{journal}{%
  \bibinfo {journal} {Phys. Rev. A}\ }%
  \textbf{\bibinfo {volume} {60}},\ \bibinfo {pages} {910} (\bibinfo {year}
  {1999}).%
  \bibAnnoteFile{NoStop}{kempe1999multiparticle}%
  
 \bibitem{coffman2000distributed}%
  \BibitemOpen
  \bibfield{author}{%
  \bibinfo {author} {\bibfnamefont{V.}~\bibnamefont{Coffman}},
  \bibinfo {author} {\bibfnamefont{J.}\ \bibnamefont{Kundu}},\ and\ \bibinfo
  {author} {\bibfnamefont{W. K.}\ \bibnamefont{Wootters}},\ }%
  \bibfield{journal}{%
  \bibinfo {journal} {Phys. Rev. A}\ }%
  \textbf{\bibinfo {volume} {61}},\ \bibinfo {pages} {052306} (\bibinfo {year}
  {2000}).%
 \bibAnnoteFile{NoStop}{coffman2000distributed}%
 
\bibitem{bengtsson2017geometry}%
  \BibitemOpen
  \bibfield{author}{%
  \bibinfo {author} {\bibfnamefont{K.}~\bibnamefont{{\.Z}yczkowski}}\ and\
  \bibinfo {author} {\bibfnamefont{I.}~\bibnamefont{Bengtsson}},\ }%
  \bibinfo {title} {\textit{Geometry of quantum states: an introduction to quantum entanglement} (Cambridge university press, 2017)}.%
  \bibAnnoteFile{NoStop}{bengtsson2017geometry}%
  
\bibitem{liao1954topology}%
  \BibitemOpen
  \bibfield{author}{%
  \bibinfo {author} {\bibfnamefont{S. D.}~\bibnamefont{Liao}},\ }%
  \bibfield{journal}{%
  \bibinfo {journal} {Trans. Amer. Math. Soc.}\ }%
  \textbf{\bibinfo {volume} {77}},\ \bibinfo {pages} {520} (\bibinfo {year}
  {1954}).%
  \bibAnnoteFile{NoStop}{liao1954topology}%
    
\bibitem{bhatia1983space}%
  \BibitemOpen
  \bibfield{author}{%
  \bibinfo {author} {\bibfnamefont{R.}~\bibnamefont{Bhatia}}\ and\ 
  \bibinfo {author} {\bibfnamefont{K. K.}\ \bibnamefont{Mukherjea}},\ }%
  \bibfield{journal}{%
  \bibinfo {journal} {Linear Alg. Appl.}\ }%
  \textbf{\bibinfo {volume} {52}},\ \bibinfo {pages} {765} (\bibinfo {year}
  {1983}).%
  \bibAnnoteFile{NoStop}{bhatia1983space}%
 
\bibitem{acin2000generalized}%
  \BibitemOpen
  \bibfield{author}{%
  \bibinfo {author} {\bibfnamefont{A.}~\bibnamefont{Ac{\'\i}n}},
  \bibinfo {author} {\bibfnamefont{A.}\ \bibnamefont{Andrianov}},
  \bibinfo {author} {\bibfnamefont{L.}\ \bibnamefont{Costa}},
  \bibinfo {author} {\bibfnamefont{E.}\ \bibnamefont{Jan{\'e}}},
  \bibinfo {author} {\bibfnamefont{J. I.}\ \bibnamefont{Latorre}},\ and\ 
  \bibinfo {author} {\bibfnamefont{R.}\ \bibnamefont{Tarrach}},\ }%
  \bibfield{journal}{%
  \bibinfo {journal} {Phys. Rev. Lett.}\ }%
  \textbf{\bibinfo {volume} {85}},\ \bibinfo {pages} {1560} (\bibinfo {year}
  {2000}).%
  \bibAnnoteFile{NoStop}{acin2000generalized}%

\bibitem{gelfand2008discriminants}%
  \BibitemOpen
  \bibfield{author}{%
  \bibinfo {author} {\bibfnamefont{I. M.}~\bibnamefont{Gelfand}},
  \bibinfo {author} {\bibfnamefont{M.}~\bibnamefont{Kapranov}}\ and\
  \bibinfo {author} {\bibfnamefont{A.}~\bibnamefont{Zelevinsky}},\ }%
  \bibinfo {title} {\textit{Discriminants, resultants, and multidimensional determinants} (Springer Science \& Business Media, 2008)}.%
  \bibAnnoteFile{NoStop}{gelfand2008discriminants}%
  
\bibitem{mandilara2014entanglement}%
  \BibitemOpen
  \bibfield{author}{%
  \bibinfo {author} {\bibfnamefont{A.}~\bibnamefont{Mandilara}},
  \bibinfo {author} {\bibfnamefont{T.}\ \bibnamefont{Coudreau}},
  \bibinfo {author} {\bibfnamefont{A.}\ \bibnamefont{Keller}},\ and\ 
  \bibinfo {author} {\bibfnamefont{P.}\ \bibnamefont{Milman}},\ }%
  \bibfield{journal}{%
  \bibinfo {journal} {Phys. Rev. A}\ }%
  \textbf{\bibinfo {volume} {90}},\ \bibinfo {pages} {050302} (\bibinfo {year}
  {2014}).%
  \bibAnnoteFile{NoStop}{mandilara2014entanglement}%
  
\bibitem{dur2000three}%
  \BibitemOpen
  \bibfield{author}{%
  \bibinfo {author} {\bibfnamefont{W.}~\bibnamefont{D{\"u}r}},
  \bibinfo {author} {\bibfnamefont{G.}\ \bibnamefont{Vidal}},\ and\ 
  \bibinfo {author} {\bibfnamefont{J. I.}\ \bibnamefont{Cirac}},\ }%
  \bibfield{journal}{%
  \bibinfo {journal} {Phys. Rev. A}\ }%
  \textbf{\bibinfo {volume} {62}},\ \bibinfo {pages} {062314} (\bibinfo {year}
  {2000}).%
  \bibAnnoteFile{NoStop}{dur2000three}%
  
\bibitem{luque2003polynomial}%
  \BibitemOpen
  \bibfield{author}{%
  \bibinfo {author} {\bibfnamefont{J. G.}~\bibnamefont{Luque}} and\ 
  \bibinfo {author} {\bibfnamefont{J. Y.}\ \bibnamefont{Thibon}},\ }%
  \bibfield{journal}{%
  \bibinfo {journal} {Phys. Rev. A}\ }%
  \textbf{\bibinfo {volume} {67}},\ \bibinfo {pages} {042303} (\bibinfo {year}
  {2003}).%
  \bibAnnoteFile{NoStop}{luque2003polynomial}%
  
\bibitem{eltschka2012multipartite}%
\BibitemOpen
\bibfield{author}{%
\bibinfo {author} {\bibfnamefont{C.}~\bibnamefont{Eltschka}},
\bibinfo {author} {\bibfnamefont{T.}\ \bibnamefont{Bastin}},
\bibinfo {author} {\bibfnamefont{A.}\ \bibnamefont{Osterloh}},\ and\ 
\bibinfo {author} {\bibfnamefont{J.}\ \bibnamefont{Siewert}},\ }%
\bibfield{journal}{%
\bibinfo {journal} {Phys. Rev. A}\ }%
\textbf{\bibinfo {volume} {85}},\ \bibinfo {pages} {022301} (\bibinfo {year}
{2012}).%
\bibAnnoteFile{NoStop}{eltschka2012multipartite}%

\bibitem{ren2008permutation}%
\BibitemOpen
\bibfield{author}{%
\bibinfo {author} {\bibfnamefont{X. J.}~\bibnamefont{Ren}},
\bibinfo {author} {\bibfnamefont{W.}\ \bibnamefont{Jiang}},
\bibinfo {author} {\bibfnamefont{X. X.}\ \bibnamefont{Zhou}},
\bibinfo {author} {\bibfnamefont{Z. W.}\ \bibnamefont{Zhou}},\ and\ 
\bibinfo {author} {\bibfnamefont{G. C.}\ \bibnamefont{Guo}},\ }%
\bibfield{journal}{%
\bibinfo {journal} {Phys. Rev. A}\ }%
\textbf{\bibinfo {volume} {78}},\ \bibinfo {pages} {012343} (\bibinfo {year}
{2008}).%
\bibAnnoteFile{NoStop}{ren2008permutation}%
  
\bibitem{verstraete2002four}%
\BibitemOpen
\bibfield{author}{%
\bibinfo {author} {\bibfnamefont{F.}~\bibnamefont{Verstraete}},
\bibinfo {author} {\bibfnamefont{J.}\ \bibnamefont{Dehaene}},
\bibinfo {author} {\bibfnamefont{B.}\ \bibnamefont{De Moor}},\ and\ 
\bibinfo {author} {\bibfnamefont{H.}\ \bibnamefont{Verschelde}},\ }%
\bibfield{journal}{%
\bibinfo {journal} {Phys. Rev. A}\ }%
\textbf{\bibinfo {volume} {65}},\ \bibinfo {pages} {052112} (\bibinfo {year}
{2002}).%
\bibAnnoteFile{NoStop}{verstraete2002four}%
  
\bibitem{briegel2001persistent}%
\BibitemOpen
\bibfield{author}{%
\bibinfo {author} {\bibfnamefont{H. J.}~\bibnamefont{Briegel}} and\ 
\bibinfo {author} {\bibfnamefont{R.}\ \bibnamefont{Raussendorf}},\ }%
\bibfield{journal}{%
\bibinfo {journal} {Phys. Rev. Lett.}\ }%
\textbf{\bibinfo {volume} {86}},\ \bibinfo {pages} {910} (\bibinfo {year}
{2001}).%
\bibAnnoteFile{NoStop}{briegel2001persistent}%

\bibitem{gour2010all}%
\BibitemOpen
\bibfield{author}{%
\bibinfo {author} {\bibfnamefont{G.}~\bibnamefont{Gour}} and\ 
\bibinfo {author} {\bibfnamefont{N.}\ \bibnamefont{Wallach}},\ }%
\bibfield{journal}{%
\bibinfo {journal} {J. Math. Phys.}\ }%
\textbf{\bibinfo {volume} {51}},\ \bibinfo {pages} {112201} (\bibinfo {year}
{2010}).%
\bibAnnoteFile{NoStop}{gour2010all}%
    
\bibitem{miyake2003classification}%
  \BibitemOpen
  \bibfield{author}{%
  \bibinfo {author} {\bibfnamefont{A.}~\bibnamefont{Miyake}},\ }%
  \bibfield{journal}{%
  \bibinfo {journal} {Phys. Rev. A}\ }%
  \textbf{\bibinfo {volume} {67}},\ \bibinfo {pages} {012108} (\bibinfo {year}
  {2003}).%
  \bibAnnoteFile{NoStop}{miyake2003classification}%
  
\bibitem{dhokovic2009polynomial}%
  \BibitemOpen
  \bibfield{author}{%
  \bibinfo {author} {\bibfnamefont{D\v Z.}~\bibnamefont{Dokovi\'c}} and\ 
  \bibinfo {author} {\bibfnamefont{A.}\ \bibnamefont{Osterloh}},\ }%
  \bibfield{journal}{%
  \bibinfo {journal} {J. Math. Phys.}\ }%
  \textbf{\bibinfo {volume} {50}},\ \bibinfo {pages} {033509} (\bibinfo {year}
  {2009}).%
  \bibAnnoteFile{NoStop}{dhokovic2009polynomial}%


  

  


\bibitem{ramachandran1986polarised}%
  \BibitemOpen
  \bibfield{author}{%
  \bibinfo {author} {\bibfnamefont{G.}~\bibnamefont{Ramachandran}} and\ 
  \bibinfo {author} {\bibfnamefont{V.}\ \bibnamefont{Ravishankar}},\ }%
  \bibfield{journal}{%
  \bibinfo {journal} {J. Phys. G}\ }%
  \textbf{\bibinfo {volume} {12}},\ \bibinfo {pages} {L143} (\bibinfo {year}
  {1986}).%
  \bibAnnoteFile{NoStop}{ramachandran1986polarised}%

\bibitem{fano1957description}%
  \BibitemOpen
  \bibfield{author}{%
  \bibinfo {author} {\bibfnamefont{U.}~\bibnamefont{Fano}},\ }%
  \bibfield{journal}{%
  \bibinfo {journal} {Rev. Mod. Phys.}\ }%
  \textbf{\bibinfo {volume} {29}},\ \bibinfo {pages} {74} (\bibinfo {year}
  {1957}).%
  \bibAnnoteFile{NoStop}{fano1957description}%

\bibitem{suma2017geometric}%
\BibitemOpen
\bibfield{author}{%
\bibinfo {author} {\bibfnamefont{S. P.}~\bibnamefont{Suma}},
\bibinfo {author} {\bibfnamefont{S.}\ \bibnamefont{Sirsi}},
\bibinfo {author} {\bibfnamefont{S.}\ \bibnamefont{Hegde}},\ and\ 
\bibinfo {author} {\bibfnamefont{K.}\ \bibnamefont{Bharath}},\ }%
\bibfield{journal}{%
\bibinfo {journal} {Phys. Rev. A}\ }%
\textbf{\bibinfo {volume} {96}},\ \bibinfo {pages} {022328} (\bibinfo {year}
{2017}).%
\bibAnnoteFile{NoStop}{suma2017geometric}%

\bibitem{serrano2019majorana}%
  \BibitemOpen
  \bibfield{author}{%
  \bibinfo {author} {\bibfnamefont{E.}~\bibnamefont{Serrano-Ens{\'a}stiga}} and\ 
  \bibinfo {author} {\bibfnamefont{D.}\ \bibnamefont{Braun}},\ }%
  \bibfield{journal}{%
  \bibinfo {journal} {arXiv preprint arXiv:1909.07740}\ }%
  (\bibinfo {year}
  {2019}).%
  \bibAnnoteFile{NoStop}{serrano2019majorana}%

\end{thebibliography}
\end{document}